\documentclass[12pt,a4paper]{article} 
\usepackage{graphicx}
\usepackage{xcolor}
\usepackage{amssymb}

\begin{document}

\title{Approximations for the natural logarithm from solenoid-toroid correspondence}
\renewcommand{\thefootnote}{\fnsymbol{footnote}}
\author{\. Ibrahim Semiz\footnote{e-mail: ibrahim.semiz@boun.edu.tr} 
\\ \small Physics Department, Bo\u gazi\c ci University, \\ \small Bebek, \. Istanbul, Turkey}
\date{ }

\maketitle
\renewcommand{\thefootnote}{\arabic{footnote}}\setcounter{footnote}{0}
\begin{abstract}

It seems reasonable that a toroid can be thought of approximately as a solenoid bent into a circle. The correspondence of the inductances of these two objects gives an approximation for the natural logarithm in terms of the average of two numbers. Different ways of averaging give different approximants. They are expressions simpler than Taylor polynomials, and are meaningful over a wider domain.

\end{abstract}

\section{Introduction: Motivation and Derivation}

The calculation of the inductance of an ideal solenoid,  
\begin{equation}
L_{\rm sol} = \mu_{0}\frac{N^{2}}{l} A, \label{L_solenoid}
\end{equation}
is a standard part of any introductory level college physics course. Here, $L$ denotes inductance, $\mu_{0}$ is the permeability of free space, $N$ the number of turns of the solenoid, $l$ its length and $A$ its cross-sectional area. An {\it ideal} solenoid is one that is infinitely long, i.e. we are assuming that $l >> a$, where $a$ is any length characterizing the cross-sectional area.

\begin{figure}[h!]
\caption{Geometry of a rectangular toroid. Adapted from ref.~\cite{serway}. } 
\centering
\includegraphics[width=\columnwidth]{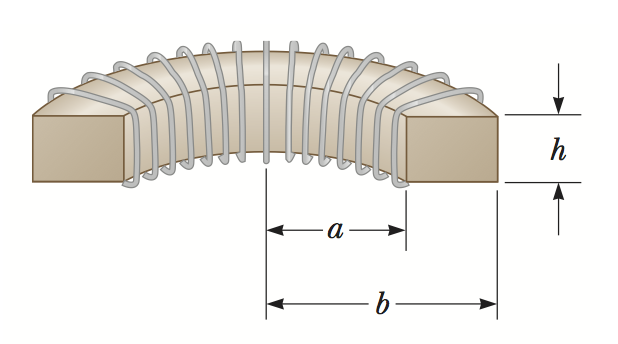}
\label{toroid} 
\end{figure}

If the said course is calculus-based, the calculation of the inductance  of a toroid with rectangular cross-section (Fig.\ref{toroid}) will often be among the end-of-chapter problems, see e.g.~\cite{serway}. The result of that calculation is
\begin{equation}
L_{\rm tor} = \mu_{0}\frac{N^{2}h}{2\pi} \ln\left(\frac{b}{a} \right), \label{L_toroid}
\end{equation} 
where $\mu_{0}$ and $N$ are as above, $a$ and $b$ are the inner and outer radii, respectively, of the toroid, and $h$ its height. 

One might wonder, in fact, an occasional student will ask; if a toroid cannot be considered as a solenoid bent into a circle. Of course, this must be a good approximation at least in {\it some} limit; and comparing (\ref{L_solenoid}) and (\ref{L_toroid}) tells us that putting $L_{\rm sol} \approx L_{\rm tor}$ will give us an approximation for the natural logarithm. We will get
\begin{equation}
\frac{h}{2\pi} \ln\left(\frac{b}{a} \right) \approx \frac{A}{l} \label{comparison}
\end{equation} 
Now, the area of the solenoid bent into the circle is $(b-a)h$, and its length is $2 \pi r$, where $r$ is some kind of average of $a$ and $b$, which we denote by $<a,b>$. Defining $b = a x$ and using the necessary property of linearity of any definition of average, we have
\begin{equation}
\ln x \approx \frac{x-1}{<x,1>}. \label{approx_main}
\end{equation} 

This is our main result. Any reasonable definition of average will give a particular approximation. In the next section, we will give three such special cases. 

\section{Particular Approximations}

We start with the most familiar concept of average, the arithmetic one, \mbox{$<a,b>_{A} \; = (a+b)/2$}. This gives
\begin{equation}
\ln x \stackrel{A}{\approx} 2 \frac{x-1}{x+1}. \label{approx_artm}
\end{equation} 
The geometric average \mbox{$<a,b>_{G} \; = \sqrt{ab}$} gives
\begin{equation}
\ln x \stackrel{G}{\approx} \frac{(x-1)}{\sqrt{x}}, \label{approx_geom}
\end{equation} 
and the harmonic average
\begin{equation}
<a,b>_{H} \; = \frac{2ab}{a+b}  \label{harm_avg}
\end{equation}
results in
\begin{equation}
\ln x \stackrel{H}{\approx} \frac{(x^{2}-1)}{2x}. \label{approx_harm}
\end{equation} 

While the approximation (\ref{approx_geom}) is not a rational function due to the presence of the factor $\sqrt{x}$, the approximations (\ref{approx_artm}) and (\ref{approx_harm}) are; i.e. they are fractions of polynomials. In fact, when aproximating a function, such an expression is called a {\em Pad\'{e} approximant}~\cite{pade}.

Another common type of average is the weighted average, but no clear motivation exists for weighting the inner and outer radius of the toroid differently, nor are there any guidelines for what the weighting factors would be; so we do not use this average at this point. On the other hand, the harmonic average (\ref{harm_avg}) can be seen as a special weighted average, where each number is weighted by the other.

\section{Comparisons}

\begin{figure}[h!]
\caption{Graphs of the function $\ln x$ (red), the first 13  Taylor polynomials of that function around $x_{0}=1$, and the three approximants (blue) of this work. The $x$-range is about five units.} 
\centering
\includegraphics[width=\columnwidth]{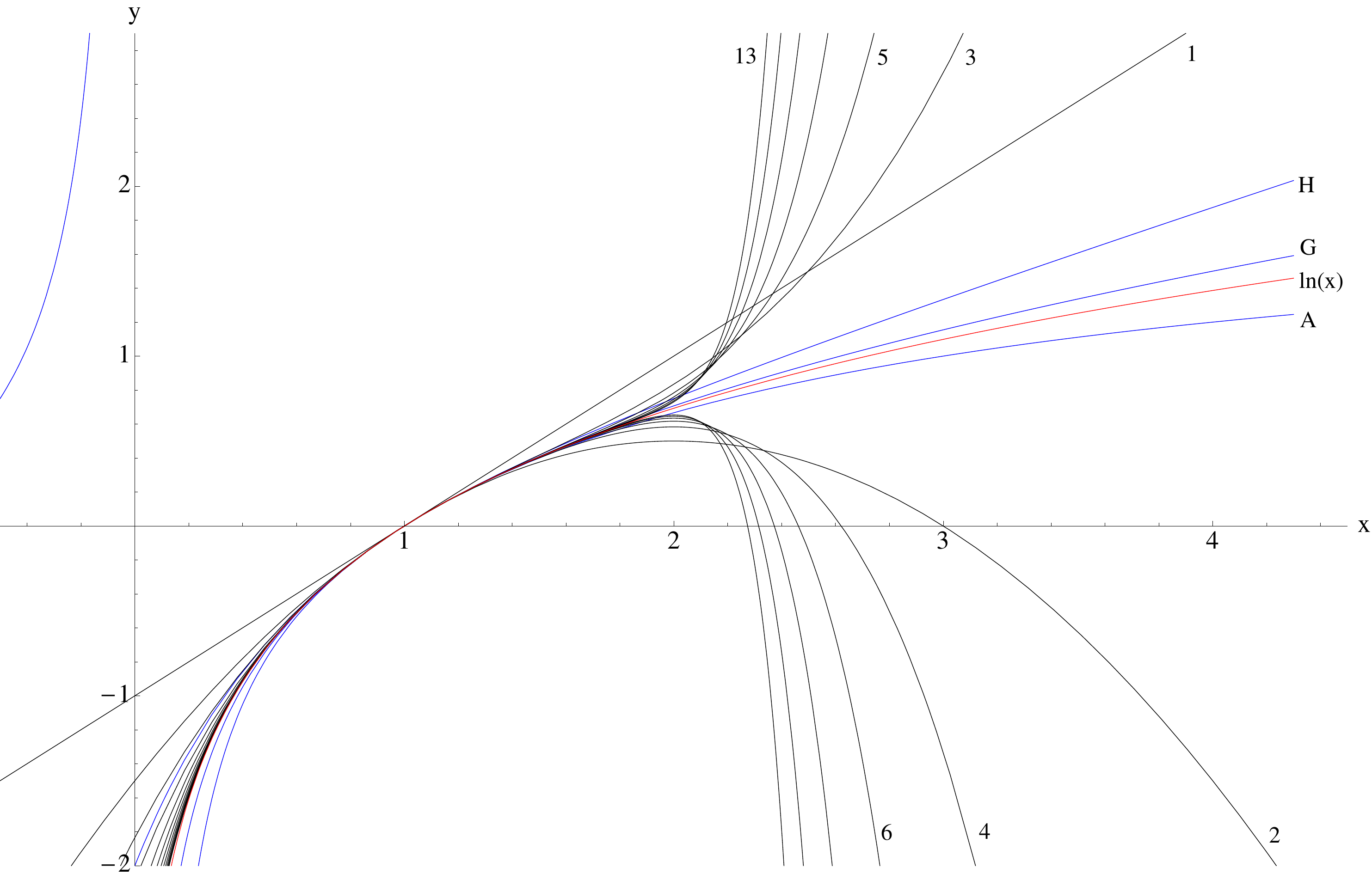}
\label{all_init} 
\end{figure}

If an approximation is needed for a function, the immediate impulse, almost reflex, of a physicist is to use a Taylor series. However, when we construct the Taylor series of the function we are interested in here, $f(x)=\ln x$, around $x_{0}=1$ (the point $x_{0}=1$ is dictated by our problem: $x>1$, since $b>a$), and make plots of the Taylor polynomials (truncated Taylor series, henceforth occasionally abbreviated as TP's), we see that at $x$-values beyond 2.5, the polynomials are totally useless (Fig.\ref{all_init}); in fact, the higher the number of terms taken, the worse a representation of the function the series is. On the other hand, our approximations (\ref{approx_artm}), (\ref{approx_geom}) and (\ref{approx_harm}) are also shown on that figure, and they perform much better in that range.

\begin{figure}[h!]
\caption{Graphs of the function $\sin x$ (red), the first 7  Taylor polynomials of that function around $x_{0}=0$. The Taylor polynomials are labeled by their orders.} 
\centering
\includegraphics[width=\columnwidth]{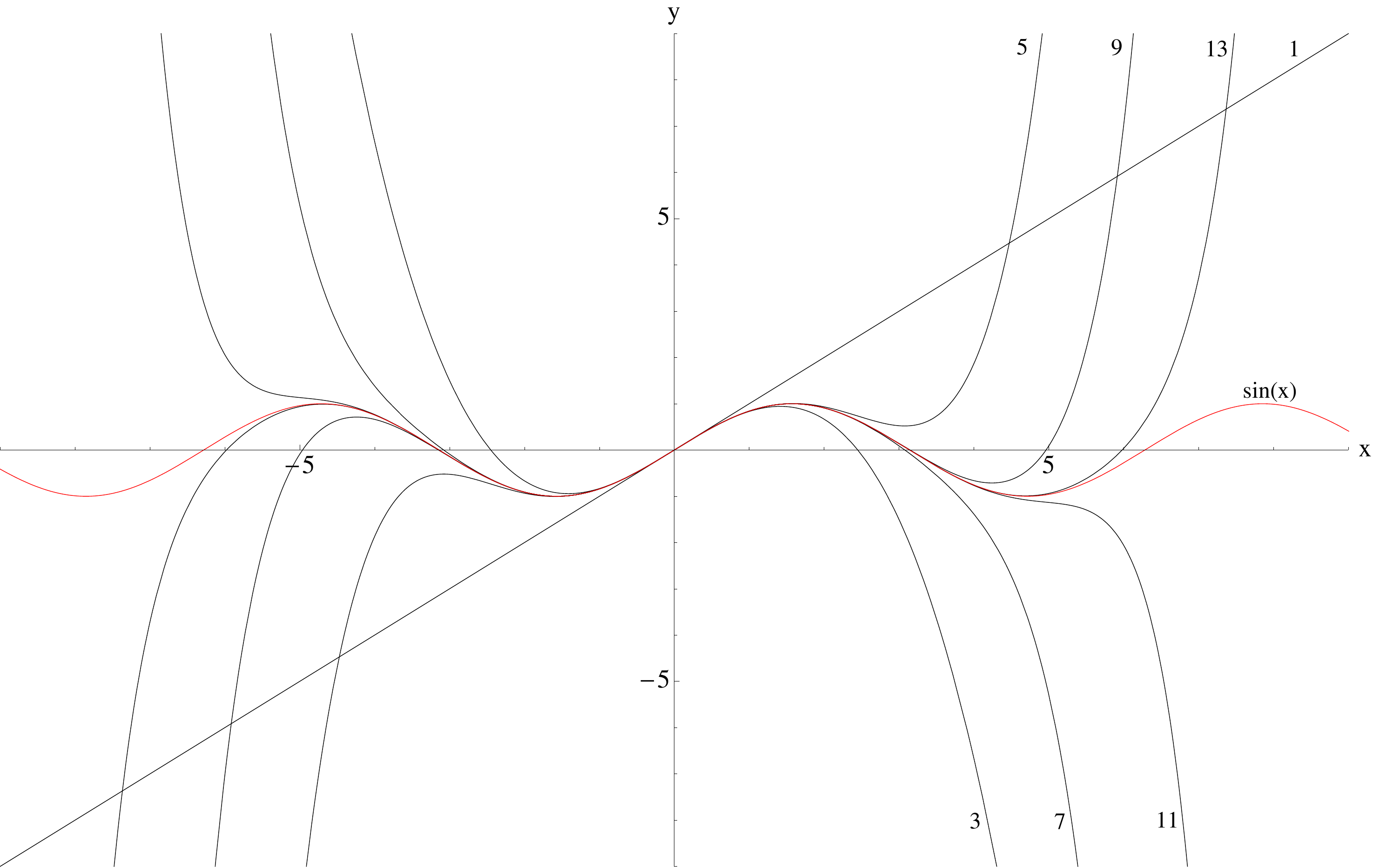}
\label{sin} 
\end{figure}

Contrast this behavior of the Taylor polynomials with the corresponding case for the function $f(x) = \sin x$ (Fig.\ref{sin}): Here, the higher the order of the polynomial, the later it peels off from the curve of $f(x)$, i.e. by increasing the order of the polynomial, we can find one that will be a good approximation in any desired range around the origin.
The difference comes from the convergence properties of the respective Taylor series: The series for $\sin x$ converges for all $x$, while for $\ln x$, around $x_{0}=1$, the radius of convergence $R$ is 1, as can be shown with standard techniques (e.g. \cite{calculus}). In fact, an upper limit of 1 on $R$ could have been guessed without calculation, by noting the singularity of the function at $x=0$.

\begin{figure}[h!]
\caption{Graphs of the function $\ln x$ (red), the first 13  Taylor polynomials of that function around $x_{0}=1$, and the three approximants (blue) of this work, shown near $x=2$.} 
\centering
\includegraphics[width=\columnwidth]{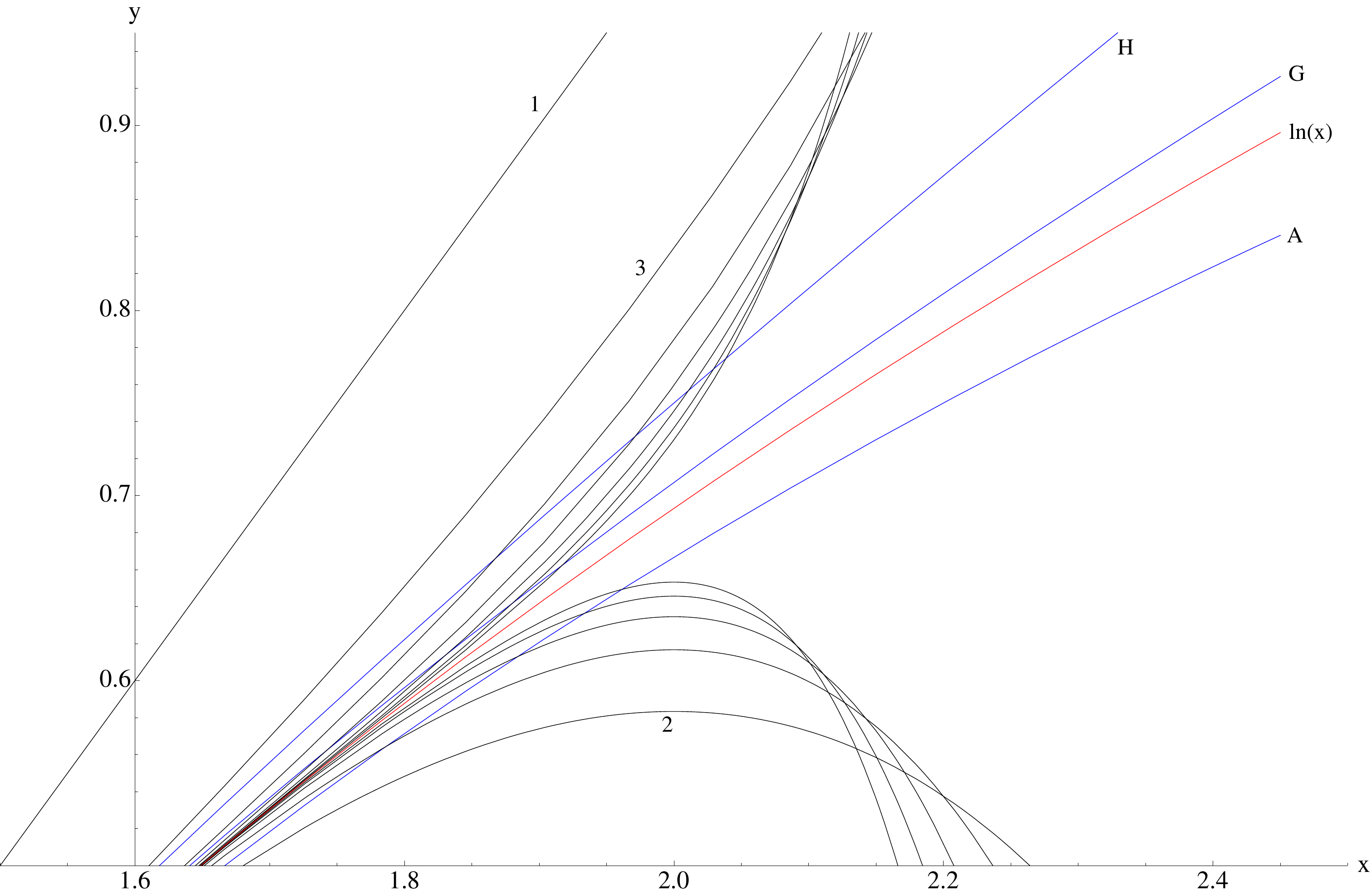}
\label{near2} 
\end{figure}

This tells us that for this particular function, the Taylor series expanded around $x_{0}=1$ is meaningless for $x \geq 2$, hence, other approximations are needed. While Taylor series around other $x_{0}$ values can be constructed, they necessitate calculation of $\ln x_{0}$ first; the approximation scheme (\ref{approx_main}) provides a neat alternative. Comparing to the TP's  around $x_{0}=1$; already at $x=1.95$, the geometric approximant is better than the first 13 TP's, and at $x=2.05$, now sounding very naturally, all three realizations of (\ref{approx_main}) are (Fig.\ref{near2}). The error of these approximants increases monotonically with $x$, since the solenoid ($\equiv$ constant magnetic field inside) approximation for the toroid becomes monotonically worse with $x$; 
the error of the geometric approximant is \%2 at $x=2$, \%5 at $x=3$, \%11 at  $x=5$ and \%24 at  $x=10$, still usable for some purposes. At large $x$ values, the approximants tend to 2, $\sqrt{x}$ and $x/2$, respectively, and become useless.

\begin{figure}[h!]
\caption{Graphs of the function $\ln x$ (red), the first 13  Taylor polynomials of that function around $x_{0}=1$, and the three approximants (blue) of this work, shown near $x=0.4$.} 
\centering
\includegraphics[width=\columnwidth]{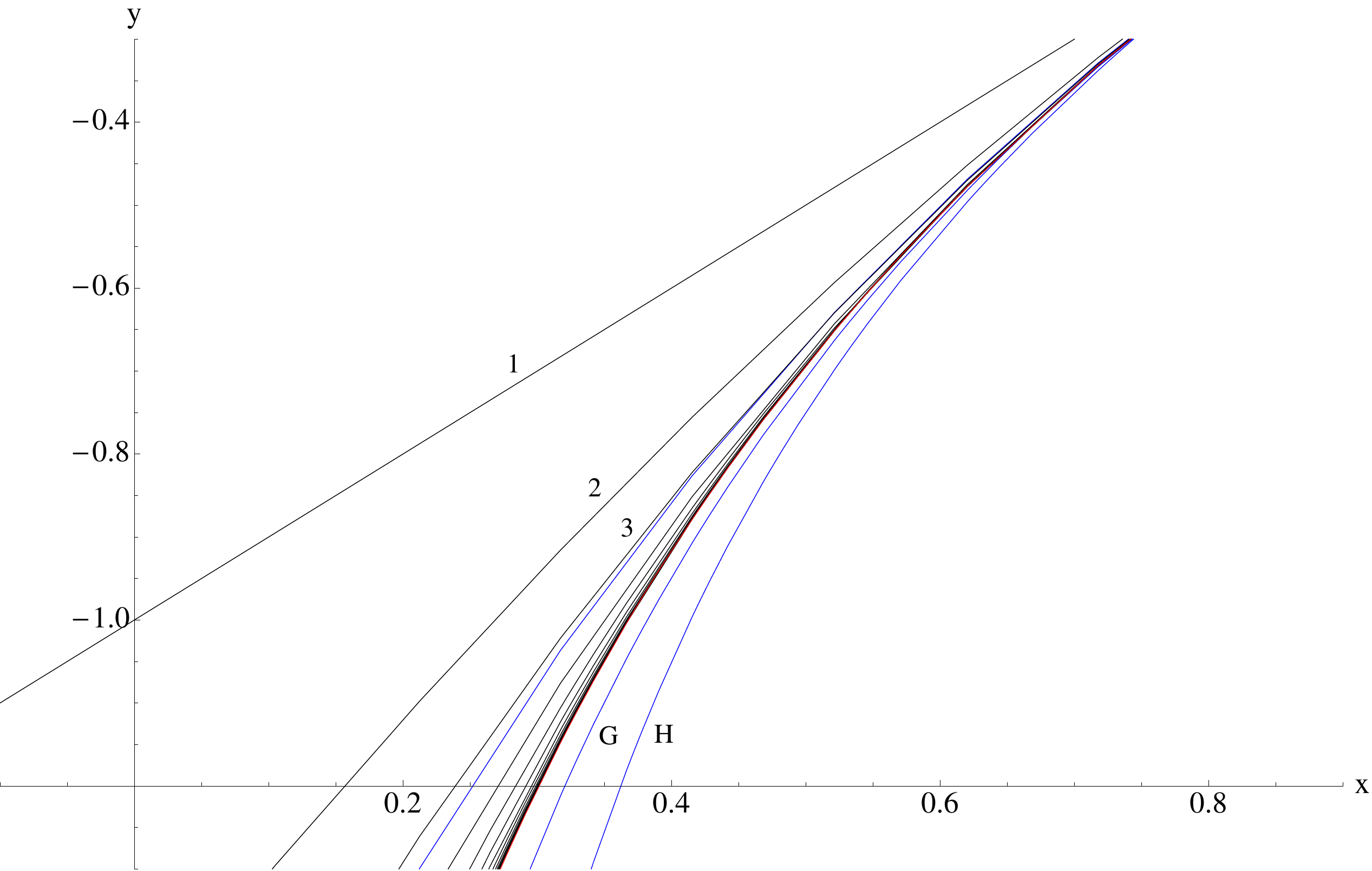}
\label{near05} 
\end{figure}

While the motivation for the scheme (\ref{approx_main}) has $x>1$ as a built-in feature, we can also check the region between $x=0$ and $x=1$. For most of this region, the fifth and higher order Taylor polynomials are better than all three of our approximants, the geometric approximant is of comparable accuracy to the fourth order TP (near $x=1/2$, TP4 is better, near $x=1/3$, the approximant), followed by the arithmetic one, the third order TP  and the harmonic approximant, in this order. The error of the geometric approximant increases to \%5 near $x=1/3$ (Fig.\ref{near05}).

Near $x=1$, that is, for $|x-1|<<1$, the Taylor poynomials are good approximations, as our reflexes tell us; in fact, the closer to $x=1$ we are, the shorter the Taylor polynomial adequate for a given level of accuracy.
However, the accuracy of all three of our approximants also increases towards perfection as we near $x=1$ from either direction, so in this region also, they are eminently usable. On the other hand, the singularity at $x=0$ cannot be exhibited by any of the Taylor polynomials, while the geometric and harmonic approximants do a visually good job of it, even if the errors are large.

One final note is that, having seen the behavior of the approximants (Figs. \ref{all_init} and \ref{near2}), some kind of average of the geometric and arithmetic averages will give a more accurate closed-expression result. For example, their geometric average,
\begin{equation}
\ln x \stackrel{4A}{\approx} (x-1) \sqrt \frac{2}{(x+1)\sqrt{x}}  \label{approx_4}
\end{equation} 
has only \%6 error at $x=10$, and the error could be decreased even more by weighting the factors. But the higher accuracy of expression (\ref{approx_4}) and analogous ones comes at the cost of losing some of the simplicity of expressions 
(\ref{approx_main}), (\ref{approx_artm}), (\ref{approx_geom}) and (\ref{approx_harm}).

\section{Conclusion}

Starting from the reasonable assumption that a toroid can in {\it some} limit be thought of as a solenoid bent into a circle, we derived the simple and neat approximation scheme (\ref{approx_main}) for the natural logarithm function. Every reasonable definition of the average of two numbers will give a particular realization of the scheme, and we exhibited three such approximants, using the most common types of average.  Among the three, the approximant based on the geometric average, (\ref{approx_geom}), is the best performer, with \%5 error at $x=1/3$ and $x=3$, the error decreasing monotonically down to zero as $x=1$ is approached from either side.   

These approximants are much simpler expressions than multi-term Taylor polynomials around $x_{0}=1$, which do not make sense for $x \gtrapprox 2$ in any case, and their accuracy is  competitive with said polynomials for the interval $0.3 < x < 2 $, as well. They could be combined for more accurate approximations, at the cost of losing some of the simplicity.

\section*{Acknowledgements}

The author thanks A. Kaz{\i}m \c{C}aml{\i}bel for help with the figures.

\end{document}